\documentclass[12pt]{iopart}

\usepackage{iopams}  
\usepackage{graphicx}
\usepackage{bm}
\usepackage{txfonts}
\usepackage{booktabs}
\usepackage{color}
\usepackage{multirow}

\newcommand{\cel}{\ensuremath{c_{\rm el}}}

\newcommand{\Dcel}{\ensuremath{\Delta c_{\rm el}}}
\newcommand{\gn}{\ensuremath{\gamma_{\rm n}}}

\newcommand{\gres}{\ensuremath{\gamma_{\rm res}}}

\newcommand{\Hc}{\ensuremath{H_{\rm c}}}
\newcommand{\Hsc}{\ensuremath{H_{\rm sc}}}
\newcommand{\Tc}{\ensuremath{T_{\rm c}}}
\newcommand{\HAC}{\ensuremath{H_{\rm ac}}}
\newcommand{\HDC}{\ensuremath{H_{\rm dc}}}
\newcommand{\Hcz}{\ensuremath{H_{\rm c2}}}

\newcommand{\cAC}{\ensuremath{\chi_{\rm ac}}}
\newcommand{\chip}{\ensuremath{\chi '}}
\newcommand{\chipp}{\ensuremath{\chi ''}}

\newcommand{\kGL}{\ensuremath{\kappa_{\rm GL}}}
\newcommand{\HT}{$H$\,--\,$T$}
\newcommand{\mOc}{\ensuremath{\muup\Omega{\rm cm}}}

\begin{document}

\title{Superconductivity in heavily boron-doped silicon carbide}

\author{M Kriener$^1$, T Muranaka$^2$, J Kato$^2$, Z A Ren$^2$, J Akimitsu$^2$ and Y Maeno$^1$}

\address{$^1$Department of Physics, Graduate School of Science, Kyoto University, Kyoto 606-8502, Japan}
\address{$^2$Department of Physics and Mathematics, Aoyama-Gakuin University, Sagamihara, Kanagawa 229-8558, Japan}

\ead{mkriener@scphys.kyoto-u.ac.jp}

\begin{abstract}
The discoveries of superconductivity in heavily boron-doped diamond (C:B) in 2004 and silicon (Si:B) in 2006 renew the interest in the superconducting state of semiconductors. Charge-carrier doping of wide-gap semiconductors leads to a metallic phase from which upon further doping superconductivity can emerge. Recently, we discovered superconductivity in a closely related system: heavily-boron doped silicon carbide (SiC:B). The sample used for that study consists of cubic and hexagonal SiC phase fractions and hence this lead to the question which of them participates in the superconductivity. Here we focus on a sample which mainly consists of hexagonal SiC without any indication for the cubic modification by means of x-ray diffraction, resistivity, and ac susceptibility.
\end{abstract}


\section{Introduction}
The possibility to achieve a superconducting phase in wide-band-gap semiconductors was suggested already in 1964 by Cohen, namely in Ge and GeSi \cite{cohen64a}. Right after the prediction several semiconductor-based compounds were indeed found to be superconducting at rather low temperatures and high charge-carrier doping concentrations. In the last decade superconductivity was found in doped silicon clathrates \cite{kawaji95a,grosche01a,connetable03a} which exhibit similar bond lengths in a covalent tetrahedral $sp^3$ network like diamond. In 2004 eventually type-II superconductivity was found in highly boron-doped diamond \cite{ekimov04a}, the cubic carbon modification with a large band gap. The boron (hole) concentration of the sample was reported to be about $n=1.8\times 10^{21}$\,cm$^{-3}$ with a critical superconducting transition temperature $\Tc \approx 4.5$\,K and an upper critical field strength $\Hcz\approx 4.2$\,T. At higher doping concentrations $n=8.4\times 10^{21}$\,cm$^{-3}$ \Tc\ was found to increase to about 11.4\,K and \Hcz\ to about 8.7\,T \cite{umezawa05a}. Subsequently its next-period neighbor in the periodic system cubic silicon was found to be a type-II superconductor in 2006 at boron concentrations of about $n=2.8\times 10^{21}$\,cm$^{-3}$ \cite{bustarret06a}. However, the critical temperature is only 0.4\,K and the upper critical field 0.4\,T \cite{commentSiB}.

\begin{figure}
\centering
\includegraphics[width=10cm,clip]{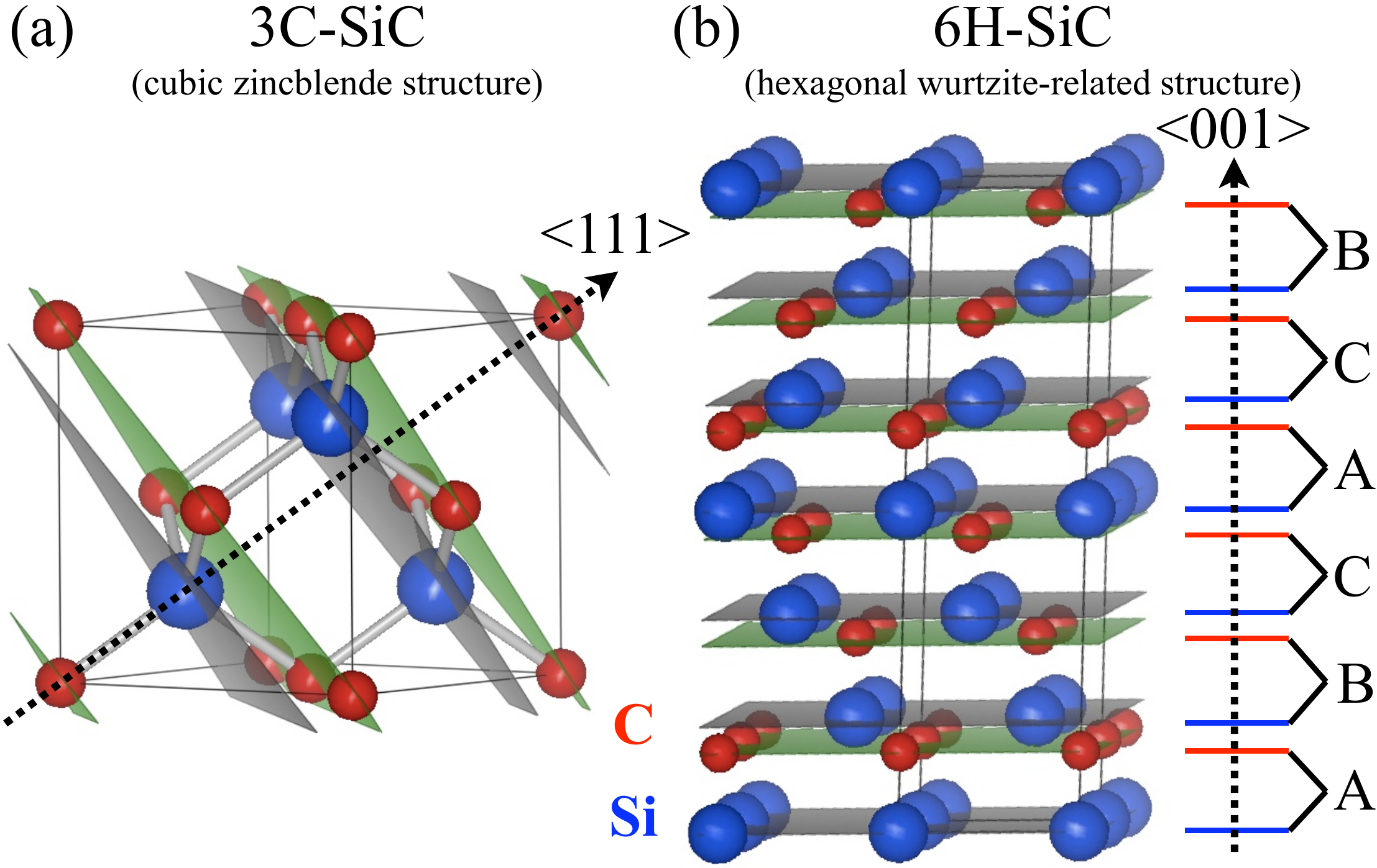}
\caption[]{(colour online) (a) Unit cell of cubic 3C-SiC. The planes mark the three C\,--\,Si bilayers forming the unit cell (stacking sequence: ABC\,--\,\dots along $\left<111\right>$ (dotted arrow)). The tetrahedral bond alignment of diamond is emphasized demonstrating the close relation to that structure. (b) Four unit cells of hexagonal 6H-SiC. The six bilayers needed for the unit cell are again denoted by planes (stacking sequence ABCACB\,--\,\dots along $\left<001\right>$ (dotted arrow)). For the drawings the software \textit{Vesta} was used \cite{momma08a}.} \label{structureSiC}
\end{figure}

In 2007 we found superconductivity in the stochiometric composition of carbon and silicon: heavily boron-doped silicon carbide \cite{ren07a}. 
One interesting difference between these three superconducting systems is the well-known polytypism in SiC which means that it exhibits different structural ground states the energy of which only slightly differs. For SiC more than 200 such structural modifications are reported \cite{casady96a}. It exists one cubic ''C'' modification labeled as 3C-SiC (zincblende = diamond structure with two different elements) or $\beta$-SiC. All other observed unit cells are either hexagonal ''H'' (wurtzite or wurtzite related) or rhombohedral ''R'', labeled as $m$H-SiC or $\alpha$-SiC and $m$R-SiC, respectively. The variable $m$ indicates the number of carbon and silicon bilayers which are needed to form the unit cell. The most important hexagonal modifications are the only pure hexagonal 2H-SiC and the of cubic and hexagonal bonds consisting 4H- and 6H-SiC polytypes. 
Fig.\,\ref{structureSiC} (from Ref.\,\cite{ren07a}) gives sketches of the two for this paper relevant modifications 3C-SiC and 6H-SiC.

For a more comprehensive introduction to SiC compare Ref.\,\cite{kriener08a} and references therein. We note here that all SiC polytypes break inversion symmetry which is known to give rise to quite unconventional superconducting scenarios, e.\,g., in heavy-fermion compounds, another principal difference to the inversion-symmetry conserving systems C:B and Si:B. However due to the comparably light elements carbon and silicon we do not believe that any exotic superconducting scenario applies to SiC:B. 

\section{Superconducting properties of SiC:B}
In the discovery paper (Ref.\,\cite{ren07a}) we used a multiphase polycrystalline boron-doped SiC sample which contains three different phase fractions: 3C-SiC, 6H-SiC, and unreacted Si. The charge-carrier concentration of this particular sample was estimated to $1.91\times 10^{21}$ holes\,/\,cm$^{3}$ \cite{ren07a}.
The critical temperature at which we observe a sharp transition in resistivity and ac susceptibility data is $\sim 1.45$\,K. The critical field strength amounts to $\Hc\approx 115$\,Oe, much lower than for the two parent compounds C:B and Si:B. A big surprise was the finding that SiC:B is a type-I superconductor as indicated by the observation of a clear hysteresis between data (resistivity and ac susceptibility) measured upon cooling from above \Tc\ to the lowest accessible temperature and a subsequent warming run in different applied external magnetic dc fields. This is in clear contrast to the reported type-II behaviour of C:B and Si:B \cite{ekimov04a,bustarret06a,commentSiB,takano07a}. Another surprise is the for a ''dirty'' doped semiconductor-based system, i.\,e., for a multiphase polycrystalline sample unexpected low residual resistivity $\rho_0$ at \Tc, which is as low as 60\,\mOc. Above \Tc\ the system features a metallic-like temperature dependence with a positive slope of d$\rho/$d$T$ in the whole temperature range up to room temperature with a residual resistivity ratio ${\rm RRR}=\rho({\rm 300\,K})/\rho_0$ of about 10. These observations are again in contrast especially to C:B which exhibits a more or less temperature independent resistivity with $\rho_0\approx 2500$\,\mOc\ and ${\rm RRR}\approx 1$. In a subsequent specific-heat study (Ref.\,\cite{kriener08a}) using the same sample we found a very small normal-state Sommerfeld parameter $\gn \approx 0.29$\,mJ/molK. Moreover, we could clearly demonstrate that SiC:B is a bulk superconductor as indicated by a specific-heat jump at about 1.45\,K coinciding with the critical temperature \Tc\ estimated from resistivity and ac susceptibility data. The jump in the specific heat is rather broad reflecting the polycrystalline multi-phase character of the sample used. However, the system comes up with a third remarkable surprise. The electronic specific heat $\cel/T$ exhibits a strict linear temperature dependence below its jump down to the lowest so-far accessed temperature $\sim 0.35$\,K and extrapolates almost identical to 0 for $T\rightarrow 0$. The jump hight is estimated to $\Dcel/\gn\Tc\approx 1$ which is only 1/3 of the BCS expectation of a weak-coupling superconductor close to the value theoretically expected for a superconducting gap with nodes \cite{hasselbach93a,nishizaki99a}. However, a strict linear temperature dependence is only expected well below \Tc, where the superconducting gap is nearly temperature independent. When approaching \Tc\ the specific heat should deviate from $\cel/T\propto T$ due to the reduction of the gap magnitude. We note here that the assumption of a BCS-like scenario with a residual contribution to the specific heat \gres, e.\,g., due to non-superconducting parts of the sample,  
yields a reasonable description of the data, too, with $\gres \approx 0.14$\,mJ/molK, compare Ref.\,\cite{kriener08a}. The jump height in this scenario is 1.48 almost fitting the BCS expectation of 1.43. In the description of the specific heat assuming a linear $\cel/T$ no residual contribution is needed, a respective fit to the data yields $\gres \approx 0$\,mJ/molK, as suggested by the almost perfect extrapolation of the data down to 0 for $T\rightarrow 0$. These results for \gres\ for the two approaches imply a superconducting volume fraction of about 100\,\% for the nodal gap scenario and about 50\,\% for the BCS-like scenario.

In a very recent publication \cite{kriener08b} we reported that the hexagonal phase fraction is superconducting and exhibits a similar linear temperature dependence of the electronic specific heat $\cel/T$ in the superconducting state and also a reduced jump height. In this paper we focus on this particular sample referred to as 6H-SiC in more detail and discuss the \HT\ phase diagram derived from ac susceptibility data. We give an evaluation of the Ginzburg-Landau (GL) parameter, too, and compare these results with those obtained for the afore mentioned ''mixed'' sample used in Refs.\,\cite{ren07a} and \cite{kriener08a} referred to as 3C/6H-SiC in this paper. 

\section{Experimental}
The details of the sample preparation of sample 3C/6H-SiC are given in Ref.\,\cite{ren07a}, sample 6H-SiC was synthesized in a similar way. The charge-carrier concentration of this sample is $0.25\times 10^{21}$ holes\,/\,cm$^{3}$ as estimated from a Hall effect measurement. The electrical resistivity was measured by a conventional four-probe technique using a commercial system (Quantum Design, PPMS). AC susceptibility measurements were performed using a mutual-inductance method in a commercial $^3$He refrigerator (Oxford Instruments, Heliox) inserted into a standard superconducting magnet with a small ac field of $\HAC = 0.01$\,Oe-rms at a frequency of 3011\,Hz in zero and several finite dc magnetic fields \HDC\ parallel to \HAC. 

\section{Results}
\begin{figure}
\centering
\includegraphics[width=7.5cm,clip]{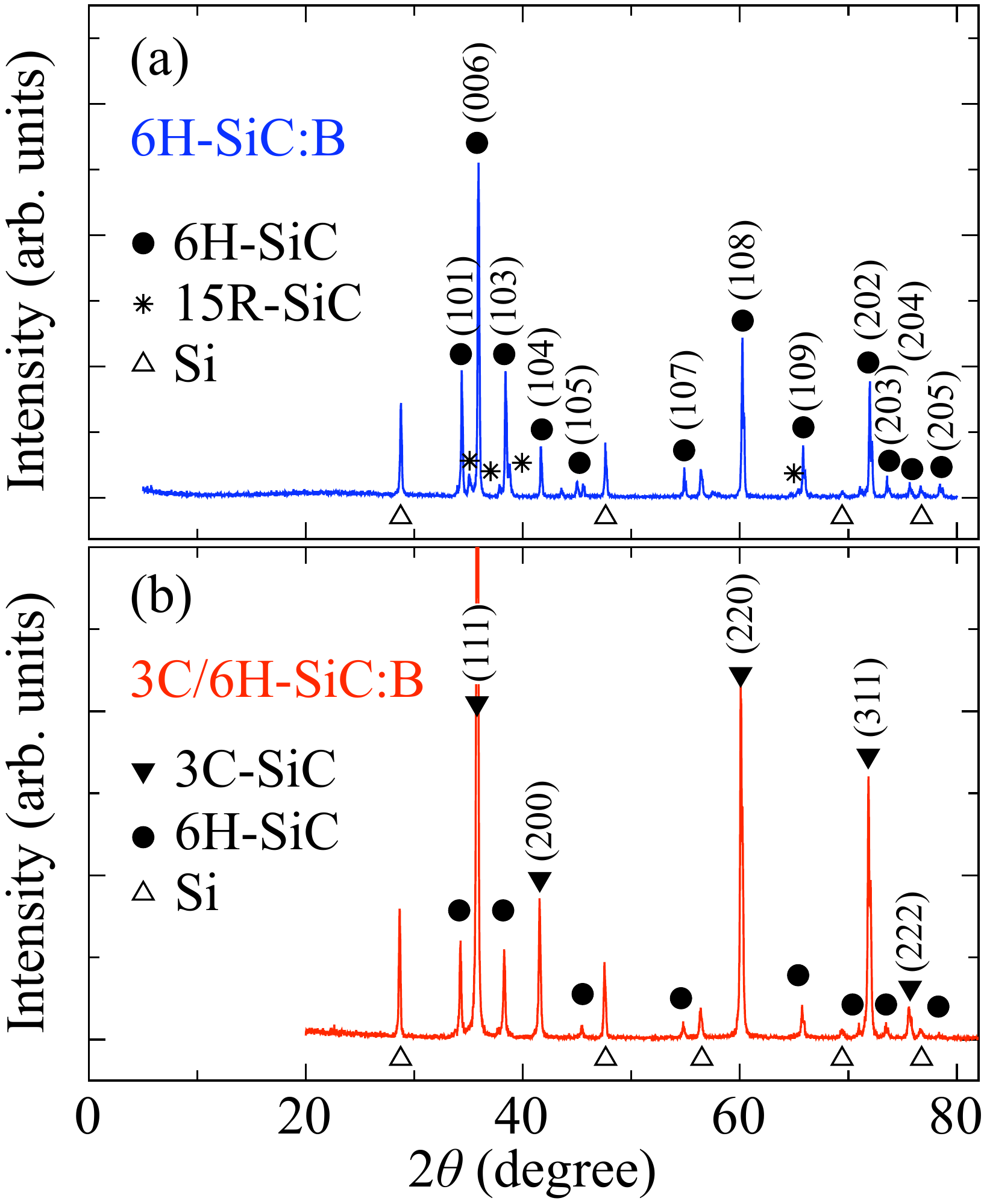}
\caption[]{(colour online) (a) Powder x-ray diffraction patterns of boron-doped 6H-SiC. Three phases, 6H-SiC, 15R-SiC, and silicon, are identified as marked by the symbols. There is no indication for a cubic SiC modification in this sample. For comparison the respective data for 3C/6H-SiC:B from Ref.\,\cite{ren07a} is shown in panel (b), too.} \label{XRay}
\end{figure}
Fig.\,\ref{XRay}\,(a) shows the result of a powder x-ray diffraction experiment of boron-doped 6H-SiC and for comparison the respective data for 3C/6H-SiC:B from Ref.\,\cite{ren07a}. The sample 6H-SiC:B is also a multiphase polycrystalline material with two different SiC modifications. However, we detect mainly hexagonal 6H-SiC and a much smaller phase fraction of rhombohedral 15R-SiC. 
In addition, we find some unreacted silicon, too. The estimation of the lattice parameters suggests substitutional boron-doping at the carbon site, compare the discussion in Ref.\,\cite{ren07a}.


\begin{figure}
\centering
\includegraphics[width=11.5cm,clip]{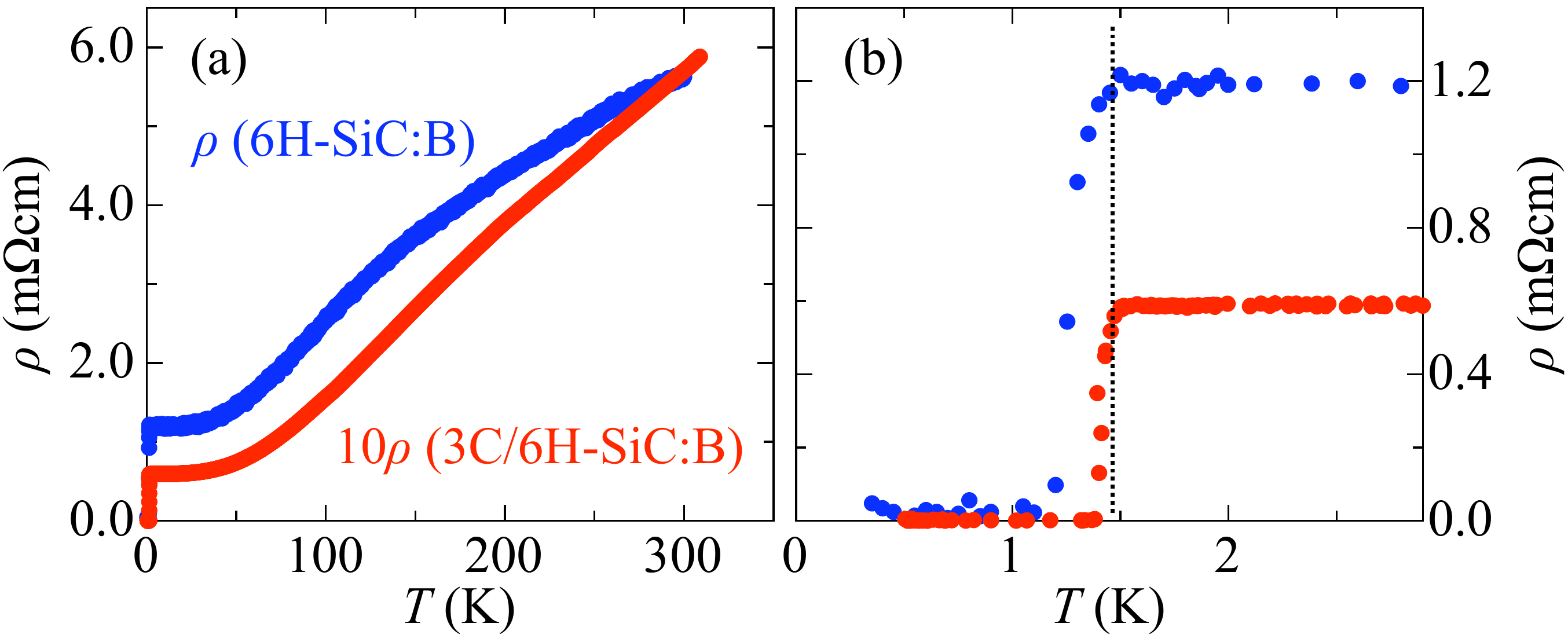}
\caption[]{(colour online) Resistivity vs.\ temperature of 6H-SiC:B (blue symbols) (a) up to room temperature and (b) around \Tc. The sample used exhibits a metallic-like temperature dependence in the whole examined temperature range above \Tc. At \Tc\ we detect a sharp drop to zero resistance. For comparison the respective data for 3C/6H-SiC:B from Ref.\,\cite{ren07a} is shown, too (red symbols). Please note that the resistivity data of 3C/6H-SiC:B is multiplied by 10. The dotted line marks the transition, see text.} \label{resistivity}
\end{figure}
In Fig.\,\ref{resistivity} the resistivity of 6H-SiC:B (blue symbols) (a) up to room temperature and (b) around the superconducting transition is presented. For comparison the respective data for 3C/6H-SiC:B (red symbols) from Ref.\,\cite{ren07a} is shown, too, here multiplied by 10. The resistivity exhibits a metallic-like behaviour, i.\,e., ${\rm d}\rho/{\rm d}T>0$, in the whole examined temperature range above \Tc. However, a linear temperature dependence is observed only in a very small temperature range above 60\,K to about 120\,K. Toward higher temperatures the slope decreases.
Below approximately 1.5\,K a clear drop to zero-resistance with a transition width of about 25\,mK is found indicating the onset of a superconducting phase. The residual resistivity is $\rho_0\approx 1.2$\,m$\Omega$cm, the residual resistivity ratio estimates to ${\rm RRR}\approx 4.7$. 

\begin{figure}
\centering
\includegraphics[width=11.5cm,clip]{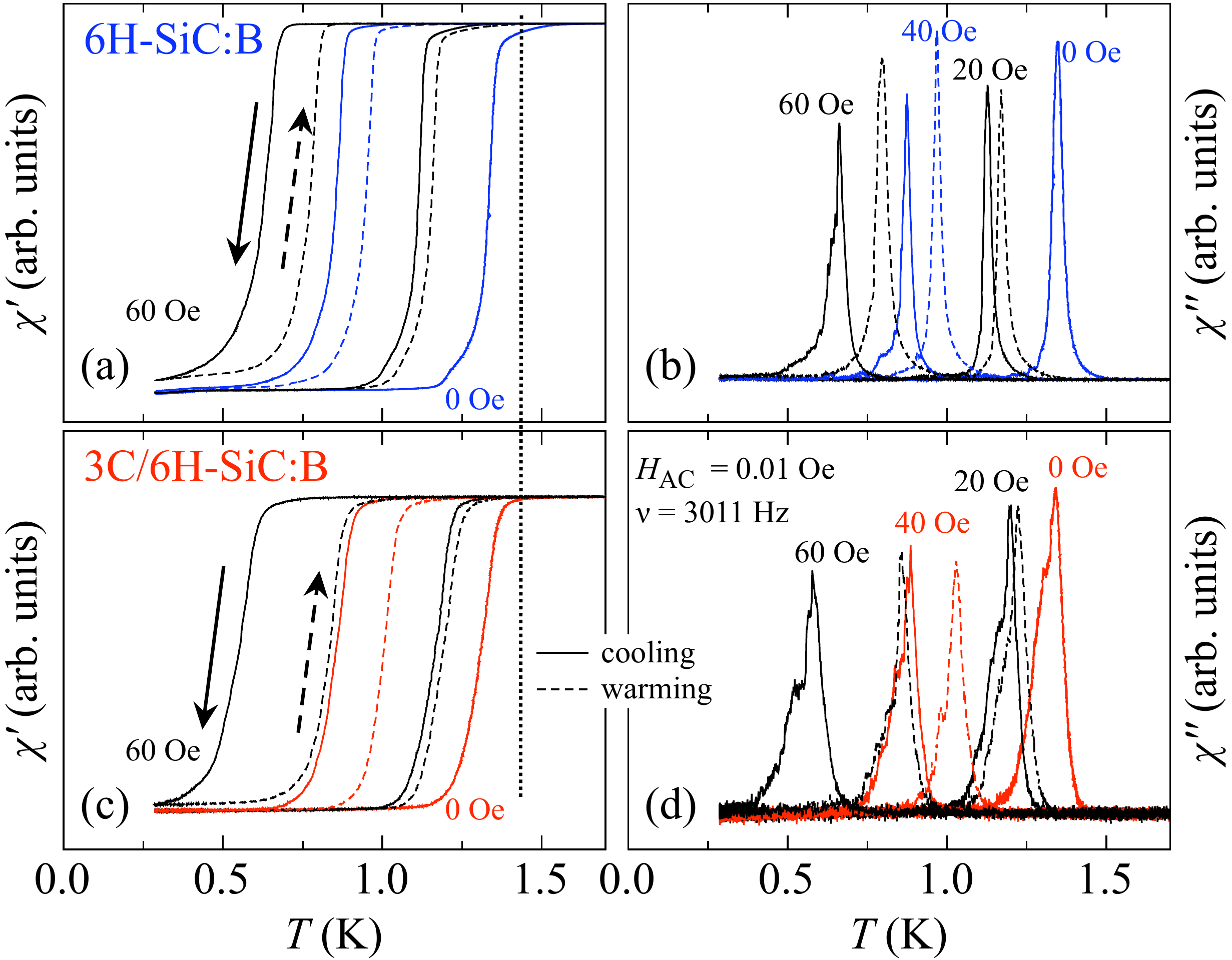}
\caption[]{(colour online) Temperature dependence of the ac susceptibility \cAC\ of 6H-SiC. In (a) the real part \chip\ and in (b) the imaginary part \chipp\ are shown. For comparison the respective data for 3C/6H-SiC:B from Ref.\,\cite{ren07a} is shown in the panels (c) and (d). The dotted line in panels (a) and (c) signals the zero-field transition temperature. The arrows in panels (a) and (c) denote the temperature sweep direction, see text.} \label{chiac}
\end{figure}
Fig.\,\ref{chiac} summarizes the results of our ac susceptibility measurements on sample 6H-SiC:B and for comparison also on the ''discovery'' sample 3C/6H-SiC:B from Ref.\,\cite{ren07a}. In panel (a) and (c) the real parts and in (b) and (d) the imaginary parts of $\cAC = \chip + i\cdot\chipp$ are shown. The measurements were performed upon cooling and warming at constant \HDC\ as well as upon sweeping \HDC\ up and down at constant temperature. The temperature dependence of the in-field susceptibility data was taken as follows: The external dc magnetic field was set above \Tc. Then the temperature was reduced down to ca.\ 300\,mK and subsequently increased above \Tc. In Fig.\,\ref{chiac} the results of temperature sweeps (solid lines: cooling run, dashed lines: warming run) for $\HDC = 0$, 20, 40, and 60\,Oe are shown. In zero field a single sharp transition with a \Tc\ of about 1.5\,K is observed in good agreement with the resistivity data. In finite magnetic fields a hysteresis between cooling and subsequent warming run appears as can be clearly seen in both, \chip\ and \chipp\ (Fig.\,\ref{chiac}\,(a) and (b)). The arrows in panel (a) denote the temperature sweep direction for $\HAC=60$\,Oe.  This in-field first-order phase transition is known as supercooling effect and is an indication of a type-I superconductor with a GL parameter $\kGL\leq 0.417$ \cite{tinkham96,feder67a,marchenko03a}. We note that we observed a similar hysteresis in field-dependent measured data at constant temperature, too. In the case of a supercooled type-I superconductor the observed critical field upon warming gives the thermodynamical critical field $\Hc(T)$. The GL parameter will be further discussed in the next section.

\begin{figure}
\centering
\includegraphics[width=11.5cm,clip]{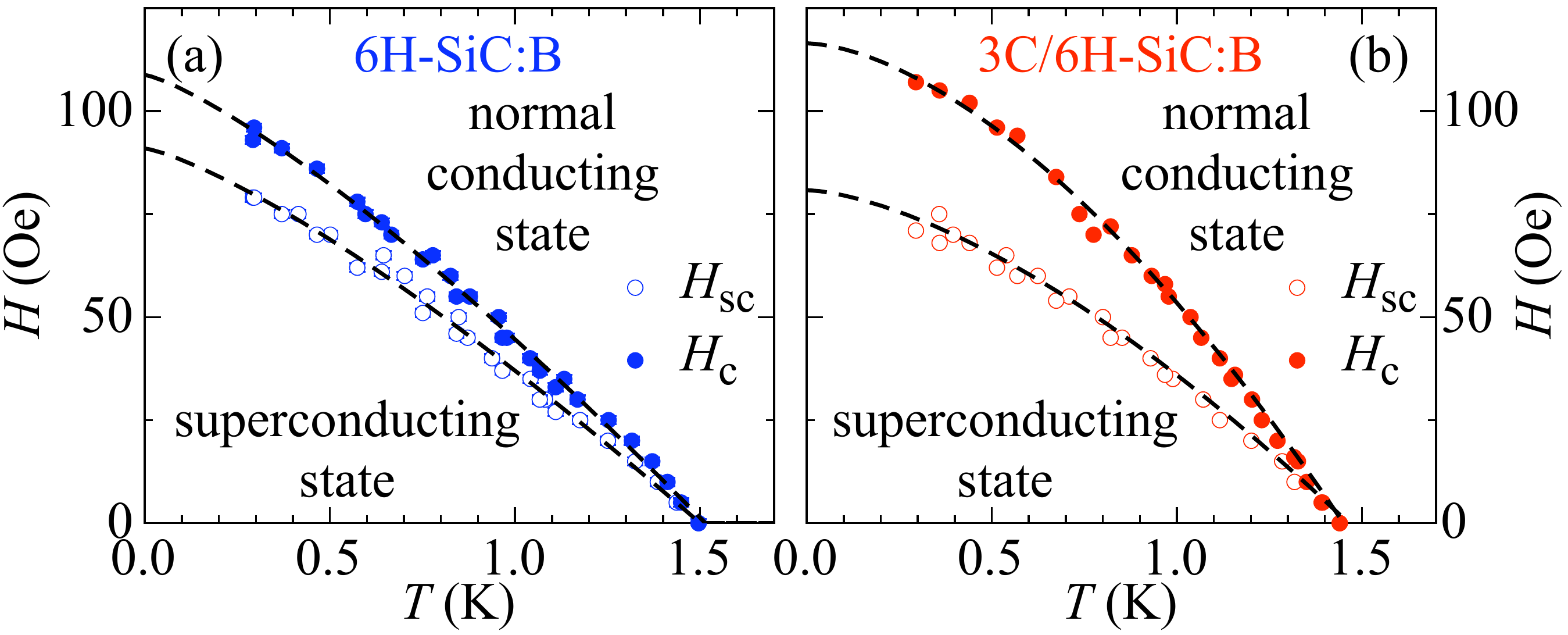}
\caption[]{(colour online) (a) \HT\ phase diagram of 6H-SiC:B. The dashed lines are fits to the data, see text. For comparison the respective data for 3C/6H-SiC:B from Ref.\,\cite{ren07a} is shown in panel (b), too. The lower phase lines \Hsc\ correspond to the ''supercooling'' effect, see text.} \label{HTdiag}
\end{figure}

From these data we derived the $H-T$ phase diagram of 6H-SiC:B shown in Fig.\,\ref{HTdiag}\,(a). For comparison the phase diagram of 3C/6H-SiC:B from Ref.\,\cite{ren07a} is shown in panel (b). The critical temperature was estimated from the shielding-fraction data \chip. Here we define \Tc\ as the temperature at which the absolute value of \chip\ has dropped by 1\,\% of the total difference in the signal between the normal and the superconducting state. This procedure was applied to both, the data obtained upon decreasing and increasing temperature. The resulting two phase lines are the lower supercooling field phase line $\Hsc(T)$ and the higher lying critical field phase line $\Hc(T)$. The black dashed lines are fits to the data applying the empirical formula $\Hc(T)=\Hc(0)[1-(T/\Tc(0))^{\alpha}]$ yielding for $T\rightarrow 0$ a supercooling field of $\Hsc(0)=(90 \pm5)$\,Oe with $\alpha=1.3$, which can be identified as the upper limit of the intrinsic supercooling limit. The thermodynamical critical field was estimated to $\Hc(0)=(110 \pm5)$\,Oe again with $\alpha=1.3$. 

\section{Discussion}
Now we would like to briefly compare these newly presented data with the data from Ref.\,\cite{ren07a} which is included to Figs.\,\ref{XRay}\,--\,\ref{HTdiag} and, together with the results of respective specific-heat studies in Refs.\,\cite{kriener08a} and \cite{kriener08b}, discuss the superconducting parameters.

The comparison of the resistivity data in Fig.\,\ref{resistivity} reveals that 6H-SiC:B ($\rho_0=1200$\,\mOc) is the much dirtier crystal as indicated by different RRR values of 5 (6H-SiC:B) and 10 (3C/6H-SiC:B). Please note, that the data of 3C/6H-SiC:B ($\rho_0=60$\,\mOc) is scaled by a factor of 10 in Fig.\,\ref{resistivity}. Obviously the transition width is smaller for the cleaner 3C/6H-SiC:B compound ($\sim 15$\,mK compared to $\sim 25$\,mK). Surprisingly the difference in the charge-carrier concentration does not affect the onset temperature of superconductivity which is for both samples almost identical around 1.5\,K, compare the dashed line in Fig.\,\ref{resistivity}. 

Like seen in Fig.\,\ref{chiac} both samples exhibit a clear supercooling behaviour which is usually observed in very clean systems like the type-I elemental superconductors. In both compounds the shielding signal is strong and robust and the shielding fraction in both cases is almost field independent, although the hysteresis is more pronounced in 3C/6H-SiC:B as seen in the \HT\ phase diagrams in Fig.\,\ref{HTdiag}. The thermodynamical critical field strength $\Hc\approx 110-115$\,Oe is again almost identical. The additional features at the low-temperature side of \chipp\ are likely due to superconducting grains with a slightly lower critical temperature. Remembering the comparably higher cleanness of sample 3C/6H-SiC:B it is surprising that these features are much less pronounced in 6H-SiC:B. Assuming that both phase fractions in 3C/6H-SiC:B are superconducting, which will be discussed next, one might speculate that the cubic phase fraction has a lower \Tc\ leading to these features. However, this conclusion is in contradiction with the finding of only one single sharp transition in the \chip\ data of that sample. 

At the time of writing Ref.\,\cite{ren07a} we were not sure which of the two existent SiC phase fractions (3C or 6H) is responsible for the superconductivity in this system. The comparison of Figs.\,\ref{XRay}\,(a) and (b) clearly demonstrates that there is no indication for a cubic phase fraction in sample 6H-SiC:B and hence the hexagonal modification of SiC:B is a bulk superconductor. Very recent data on 3C-SiC:B implies that also the cubic modification participates in the superconductivity \cite{muranaka08a}. The unreacted silicon included to our samples is likely to be an insulating phase fraction with no electronic and almost no phononic contributions to the specific heat at low temperatures. Therefore a residual contribution caused by this phase fraction cannot easily explain the values for \gres\ found in the specific-heat analysis assuming a BCS-like scenario given in Refs.\,\cite{kriener08a,kriener08b} as mentioned in the introduction. Therefore this further supports the power-law description of the data in accordance with the observation of a linear temperature dependence of $\cel/T$ below \Tc\ and a reduced jump height.

Together with the analysis of the specific-heat data we are able to estimate the superconducting parameters for 6H-SiC:B as described in detail for 3C/6H-SiC:B in Ref.\,\cite{kriener08a}. They are summarized in Table\,\ref{SiCprop}.
\begin{table}[t]
\centering
\caption{Normal-state (left) and superconducting state properties (right) of 6H-SiC:B (this work) and 3C/6H-SiC (taken from Ref.\,\cite{kriener08a}). The parameters $\ell$, $\xi(0)$, and $\lambda(0)$ are the mean-free path and the superconducting penetration depth and coherence length.}
\label{SiCprop}
\begin{tabular}{lccc}
\toprule
                   & 6H-SiC:B  & 3C/6H-SiC:B   \\ \hline \addlinespace[0.75em]
$n$ (cm$^{-3}$)    & $0.25\cdot 10^{21}$ & $1.91\cdot 10^{21}$      \\ \addlinespace[0.1em]
\gn\ (mJ/molK$^2$) & 0.349 & 0.294  \\ \addlinespace[0.1em]
$\Dcel/\gn\Tc$     & 1     & 1    \\ \addlinespace[0.1em]
$\rho_0$ (\mOc)    & 1200  & 60  \\ \addlinespace[0.1em]
RRR                & 5     & 10    \\ \addlinespace[0.1em]
$\ell$ (nm)        & 2.7   & 14  \\ \addlinespace[0.1em]
\bottomrule
\end{tabular}
\hspace{0.5cm}
\begin{tabular}{lccc}
\toprule
                   & 6H-SiC:B  & 3C/6H-SiC:B   \\ \hline \addlinespace[0.75em]
$\Tc(0)$ (K)       & 1.5   & 1.45 \\ \addlinespace[0.1em]
$\Hc(0)$ (Oe)      & 110   & 115    \\ \addlinespace[0.1em]
$\Hsc(0)$ (Oe)     & 90    &  80     \\ \addlinespace[0.1em]
$\xi(0)$ (nm)      & 78    & 360 \\ \addlinespace[0.1em]
$\lambda(0)$ (nm)  & 560   & 130   \\ \addlinespace[0.1em]
\kGL(0)            & 7     & 0.35 \\ \addlinespace[0.1em]
\bottomrule
\end{tabular}
\end{table}
Here the GL parameter is derived from the experimental values of the charge-carrier concentration $n$, the critical temperature \Tc, and the Sommerfeld parameter of the normal-state specific heat \gn. Using the values given in Table\,\ref{SiCprop} yields $\kGL =7$. This is much higher than the value suggested by the finding of a supercooling effect $\kGL \leq 0.417$ and the finding for 3C/6H-SiC:B, i.\,e., $\kGL=0.35$. The values of \Tc\ and \gn\ are similar for both samples but the charge-carrier concentration of 6H-SiC:B is about one order of magnitude smaller than that of 3C/6H-SiC:B and indeed, this is the crucial parameter. The GL parameter is defined as $\kGL=\lambda/\xi$, compare the discussion in Ref.\,\cite{kriener08a}. The penetration depth depends on the charge-carrier concentration as $\lambda\propto 1/n^{2/3}$, the coherence length as $\xi\propto n^{2/3}$ and hence $\kGL \propto 1/n^{4/3}$. Using a charge-carrier concentration of about $2\times10^{21}$/cm$^{-3}$ yields a GL parameter of the right order of magnitude. Therefore the determination of $n$ needs further clarification. 

\section{Summary}
In summary we present a study of mainly hexagonal boron-doped 6H-SiC without any indication of a cubic 3C phase fraction by means of x-ray diffraction, resistivity, and ac-susceptibility data. We compare these results with those obtained for a sample containing reasonable fractions of both, the 3C- and the 6H-SiC modification. Both samples are bulk type-I superconductors as indicated by the \HT\ phase diagrams, i.\.e., the finding of a pronounced supercooling indicated by an in-field first-order phase transition, and recent specific-heat studies, revealing a clear jump at \Tc. The sample consisting of both SiC phase fractions turns out to contain a one order of magnitude higher charge-carrier concentration. Moreover it is the much cleaner system as indicated by a surprising low residual resistivity. The GL parameter for the hexagonal SiC sample is estimated to be much higher than expected for a type-I superconductor with a strong supercooling effect. This could be due to an erroneous determination of the charge-carrier concentration and needs further clarification. Taking all data together we have strong indications that boron-doped cubic as well as hexagonal SiC are bulk superconductors.

\section*{Acknowledgements} 
This work was supported by a Grants-in-Aid for the Global COE ''The Next Generation of Physics, Spun from Universality and Emergence'' from the Ministry of Education, Culture, Sports, Science, and Technology (MEXT) of Japan, and by the 21st century COE program ''High-Tech Research Center'' Project for Private Universities: matching fund subsidy from MEXT. It has also been supported by Grants-in-Aid for Scientific Research from MEXT and from the Japan Society for the Promotion of Science (JSPS). TM is supported by Grant-in-Aid for Young Scientists (B) (No. 20740202) from MEXT and MK is financially supported as a JSPS Postdoctoral Research Fellow.

\section*{References}

\begin{thebibliography}{10}
\bibitem{commentSiB}
{A presentation during this workshop reported that new generation Si:B samples
  exhibit a higher \Tc\ of about 0.7\,K and a much reduced upper critical field
  strength of $\Hcz\approx 700$\,Oe offering the possibility that Si:B could be
  a type-I superconductor.}
\end{thebibliography}

\providecommand{\newblock}{}

\end{document}